\documentclass[a4paper,twoside]{article}

\usepackage{epsfig}
\usepackage{subcaption}
\usepackage{calc}
\usepackage{amssymb}
\usepackage{amstext}
\usepackage{amsmath}
\usepackage{amsthm}
\usepackage{multicol}
\usepackage{pslatex}
\usepackage{apalike}
\usepackage{algorithm2e}
\usepackage[bottom]{footmisc}
\usepackage{url}
 
\usepackage{SCITEPRESS}     

\begin{document}

\title{Implicit Predecessor-Based Region of Attraction Estimation and Robust Invariance Analysis for a Two-Wheeled Inverted Pendulum}

\author{
\authorname{Lorenzo Fici\sup{1}\orcidAuthor{0009-0001-9450-1441} and Guillaume Ducard\sup{1}\orcidAuthor{0000-0002-7400-4915}*}
\affiliation{\sup{1}Université Côte d‘Azur, i3S/CNRS, Sophia Antipolis, France}
\email{fici@i3s.unice.fr, ducard@i3s.unice.fr}
}


\abstract{Estimating the region of attraction (RoA) of nonlinear systems is fundamental for assessing closed-loop stability and ensuring safe operation. While Lyapunov-based approaches provide certified stability guarantees, they often yield conservative inner approximations of the RoA. This paper combines a certified Lyapunov-based positively invariant set with a predecessor-based implicit representation to compute a significantly less conservative inner approximation of the RoA while preserving formal stability guarantees. In addition, the robust positive invariance of the initial certified Lyapunov-based invariant set is analyzed under bounded additive input disturbances, providing formal robustness guarantees. The proposed methodology is demonstrated on a nonlinear two-wheeled inverted pendulum stabilized by a saturated linear quadratic regulator. The resulting RoA approximation is compared with the initial Lyapunov-certified invariant set and validated through Monte Carlo simulations and hardware experiments, showing a substantially enlarged certified operating region that matches the empirical closed-loop behavior. These results demonstrate the practical applicability of combining certified Lyapunov analysis with predecessor-based set propagation for RoA approximation and robustness assessment of nonlinear systems.}

\onecolumn \maketitle \normalsize \setcounter{footnote}{0} \vfill

\begingroup
\renewcommand{\thefootnote}{}
\footnotetext{*This work was supported by the French government through the France 2030 investment plan managed by the National Research Agency (ANR), as part of the Initiative of Excellence Université Côte d'Azur under reference number ANR-15-IDEX-01.}
\endgroup

\section{\uppercase{Introduction}}
\label{sec:introduction}
The Region of Attraction (RoA) is a fundamental concept in nonlinear control, defined as the set of all initial conditions from which the closed-loop system asymptotically converges to a desired equilibrium~\cite{khalil_nonlinear_2002}. As such, the RoA provides a measure of closed-loop stability, characterizing the range of operating conditions for which a controller can successfully recover the system. Despite its importance, computing the exact RoA of nonlinear systems is generally intractable, and practical approaches typically rely on either conservative analytical approximations or computationally intensive numerical methods~\cite{khalil_nonlinear_2002,gross_analytic_2022,tellez_estimate_2013,horibe_quantitative_2018}.

Two-Wheeled Inverted Pendulums (TWIPs) constitute a representative class of underactuated nonlinear systems for which RoA analysis is particularly relevant. Due to their inherently unstable dynamics and actuator limitations, the ability of a controller to recover the upright equilibrium depends strongly on the initial condition of the system~\cite{zhang_control_2025}. Consequently, the RoA provides a meaningful metric for assessing the closed-loop performance of balancing controllers and for quantifying the range of initial conditions from which balance can be successfully restored.

A variety of approaches have been proposed to approximate the RoA of nonlinear systems. Lyapunov-based methods provide certified inner approximations with formal stability guarantees but are often conservative, whereas numerical and sampling-based techniques generally yield larger approximations at the expense of losing rigorous certification. Predecessor-set methods offer a complementary perspective by iteratively propagating invariant regions under the system's nonlinear dynamics, enabling the computation of progressively larger certified approximations when initialized from a positively invariant set. Motivated by these observations, this work investigates the application of a predecessor-based implicit representation to enlarge a Lyapunov-certified invariant set for a nonlinear TWIP, while complementing the analysis with a robust positive invariance (RPI) study and experimental validation.
\subsection{Related work}
A broad range of methods has been proposed to estimate the RoA of nonlinear systems. Classical Lyapunov-based approaches provide certified but often conservative inner approximations; alternative approaches based on numerical simulations, Monte Carlo sampling, or experimental exploration generally produce less conservative approximations but sacrifice formal guarantees~\cite{gross_analytic_2022,tellez_estimate_2013,horibe_quantitative_2018}.
Several works have addressed this objective through Lyapunov function design. Vannelli and Vidyasagar~\cite{vannelli_maximal_nodate} introduced the concept of maximal Lyapunov functions, whose level sets converge to the exact domain of attraction. While theoretically appealing, constructing maximal Lyapunov functions for general nonlinear systems remains computationally demanding. More recently, Valmorbida and Anderson~\cite{valmorbida_region_2017} proposed the use of rational Lyapunov functions, i.e., $V(x)= \frac{p(x)}{q(x)}$, where $p(x)$ and $q(x)$ are polynomial functions, to obtain significantly less conservative RoA estimates than conventional polynomial or quadratic Lyapunov functions. In contrast, this work deliberately employs a quadratic Lyapunov function, prioritizing computational simplicity while leveraging the predecessor-expansion procedure as a complementary means to enlarge the resulting region.

Another line of research enlarges certified regions by propagating invariant sets through the system dynamics. Balint et al.~\cite{balint_methods_2006} formulate the expanded region as recursive preimages of a certified invariant set, whereas the proposed method expresses the predecessor expansion through an implicit scalar membership function obtained by composing the Lyapunov function with the nonlinear closed-loop dynamics. This representation enables exact membership evaluation without explicitly constructing the expanded set and naturally supports numerical validation and experimental certification. More recently, Serry and Liu~\cite{serry_underapproximating_2025} developed an implicit predecessor representation to compute underapproximations of the safe domain of attraction by iteratively expanding backward reachable sets while enforcing safety constraints. The predecessor computation adopted in the present work is closely related to this formulation. However, instead of considering explicit state-safety constraints, we account for the physical actuator limits by incorporating input saturation directly into the nonlinear closed-loop dynamics. The predecessor expansion is then used to enlarge an estimate of the RoA.

Related ideas have also been investigated using alternative set representations. Han et al.~\cite{han_enlarging_2016} proposed a zonotope-based framework, where a zonotope is a centrally symmetric convex polytope, that enlarges inner approximations of the region of attraction by recursively computing backward reachable sets while optimizing a feedback controller. Unlike the present work, their approach explicitly propagates zonotopic set representations and is primarily developed for linear discrete-time systems, whereas the proposed method starts from a Lyapunov-certified region and enlarges it through predecessor-based expansion for nonlinear closed-loop dynamics.

In summary, existing approaches either improve the quality of the initial Lyapunov approximation through more sophisticated Lyapunov functions or enlarge certified regions using backward reachability under additional assumptions such as linear dynamics or MPC feasibility. The present work combines a computationally inexpensive quadratic Lyapunov function with predecessor-set expansion to obtain larger certified estimates of the RoA for a nonlinear TWIP.

\subsection{Contributions}
The scientific contributions of this paper can be summarized as follows:
\begin{enumerate}
    \item a predecessor-based implicit representation that enlarges a certified Lyapunov-based invariant set, yielding a significantly less conservative certified inner approximation of the RoA of the upright equilibrium of the nonlinear closed-loop TWIP system than that provided by the original Lyapunov-based invariant set,
    \item a robust positive invariance analysis of the initial certified Lyapunov-based invariant set under bounded additive input disturbances.
\end{enumerate}
The remainder of this paper is structured as follows: Section~\ref{sec:nonlinear_model} presents the nonlinear model of the TWIP and the design of the LQR control policy. Section~\ref{sec:RoA_estimation} explains the proposed RoA estimation methodology. Section~\ref{sec:numerical_results} reports the numerical results obtained for the proposed approach and validates its effectiveness on the TWIP system. Section~\ref{sec:conclutions} concludes the paper and suggests directions for future research. 

\section{\uppercase{System modeling and Control Design}}
\label{sec:nonlinear_model}
\begin{figure}
    \centering
    \includegraphics[width=\linewidth]{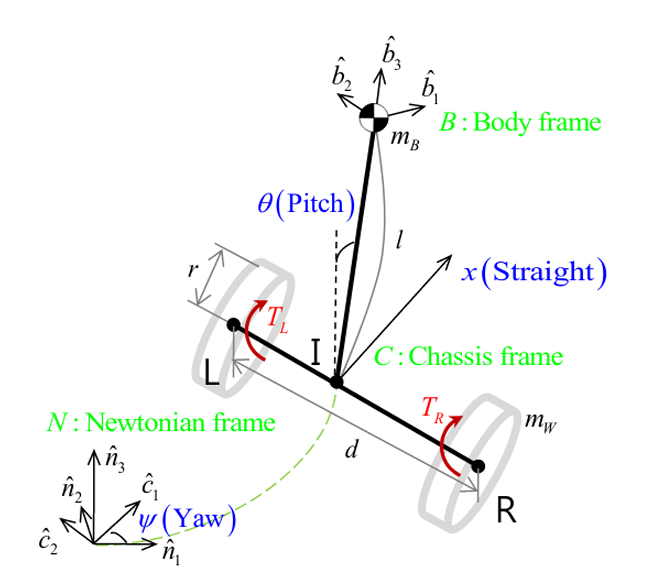}
    \caption{Schematic of the TWIP showing the coordinate frames and physical parameters. The corresponding model parameters are listed in Table~\ref{tab:model_parameters}.}
    \label{fig:sigi_schematic}
\end{figure}

This section presents the nonlinear mathematical model of the TWIP robot. The model comprises two mechanical degrees of freedom and assumes longitudinal motion only, neglecting yaw dynamics. A schematic of the robot is shown in Fig.~\ref{fig:sigi_schematic}, illustrating its main components, coordinate definitions, and physical parameters. The control objective is to stabilize the robot about its upright equilibrium configuration, corresponding to the robot balancing configuration with zero translational and angular velocities.

\subsection{Nonlinear TWIP model}
The nonlinear model, presented in~\cite{fici_comparative_2025} , serves as the basis for the computation of the Lyapunov-based invariant set, the predecessor-based expansion, the RPI analysis, and the closed-loop simulations. The equations of motion are expressed as
\begin{equation}
\label{eq:2DOF_dynamics}
\scalebox{0.93}{$
\begin{aligned}
\ddot{x}_w &= \frac{1}{d_1}
\Big( aI_O \dot{\theta}^2 \sin\theta
- a^2 g \sin\theta \cos\theta
+ T \big( \frac{I_O}{r} + a\cos\theta \big) \Big), \\[4pt]
\ddot{\theta} &= \frac{1}{d_1}
\Big( -a^2 \dot{\theta}^2 \sin\theta \cos\theta
+ a\, m_O g \sin\theta  \\[-1pt]
&\qquad 
- T\big( m_O + \frac{a}{r}\cos\theta \big) \Big), \\[4pt]
T &= 2 \, i_{gb}\frac{K}{R_M}
\Big( \text{db}_{u_0}(u_c) - K\, i_{gb}
\big( \frac{\dot{x}_w}{r} - \dot{\theta} \big) \Big),
\end{aligned}
$}
\end{equation}
\noindent where $T$ denotes the torque and the auxiliary variables are given by:
\begin{align*}
a &= m_B l, &
I_O &= I_2 + m_B l^2, \\
m_O &= m_B + 2m_W + 2\frac{J}{r^2}, &
d_1 &= I_O m_O - a^2 \cos^2 \theta.
\end{align*}
The DC motors exhibit a dead-zone $\text{db}_{u_0}(u_c)$ due to static friction and gearbox losses. To compensate for this effect, the voltage applied to the motors is augmented by a constant offset $u_0$ having the same sign as the commanded input. Consequently, $u_c$ denotes the dead-zone compensated motor voltage.
The model parameters are listed in Table~\ref{tab:model_parameters}.
\begin{table}[h]
\caption{Model parameters of the TWIP robot.}
\label{tab:model_parameters}
\centering
\begin{tabular}{|p{2.7cm}|c|c|c|}
\hline
Parameters & Sym. & Value & Unit \\
\hline
Wheel spacing & $d$ & 0.10 & m \\
\hline
Axle--COM length & $l$ & $2.75e-2$ & m \\
\hline
Wheel radius & $r$ & 0.04 & m \\
\hline
Body mass & $m_B$ & 0.368 & kg \\
\hline
Wheel mass & $m_W$ & 0.02 & kg \\
\hline
Wheel inertia & $J$ & $2.25e-5$ & kg$\cdot$m$^{2}$ \\
\hline
Gravity & $g$ & 9.81 & m/s$^{2}$ \\
\hline
Gearbox ratio & $i_{gb}$ & 49.86 & 1 \\
\hline
Motor constant & $K$ & $1.5e-3$ & N$\cdot$m/A \\
\hline
Motor resistance & $R_M$ & 12 & $\Omega$ \\
\hline
Pitch-axis inertia & $I_2$ & $2.17e-4$ & kg$\cdot$m$^{2}$ \\
\hline
$\text{db}_{u_0}$ compensation & $u_0$ & 0.625 & V \\
\hline
\end{tabular}
\end{table}

For the synthesis of the LQR controller, the nonlinear model (\ref{eq:2DOF_dynamics}) is linearized at the upright equilibrium point and discretized with exact discretization using a sampling time of $T_s = 10^{-2}\,s$, providing the following discrete-time linear dynamics
\begin{equation}
\begin{aligned}
&x_{k+1} = A_d\, x_k + B_d\, u_k, \quad
x_k = \begin{bmatrix} x_w & \dot{x}_w & \theta & \dot{\theta} \end{bmatrix}^\top, \\[3pt]
&A_d = 
\left[
\begin{array}{r@{\hspace{10pt}}r@{\hspace{10pt}}r@{\hspace{10pt}}r}
1 & 9.883e{-}3 & -6.524e{-}5 & 4.471e{-}6 \\
0 & 9.768e{-}1 & -1.276e{-}2 & 8.620e{-}4 \\
0 & 6.175e{-}3 & 1.008e{+}0 & 9.780e{-}3 \\
0 & 1.222e{+}0 & 1.573e{+}0 & 9.591e{-}1
\end{array}\right],\\[3pt]
&B_d =
\left[
\begin{array}{r}
 6.272e{-}5 \\
 1.240e{-}2 \\
-3.303e{-}3 \\
-6.533e{-}1
\end{array}
\right].
\end{aligned}
\label{eq:AB_matrices}
\end{equation}

\subsection{Saturated LQR}
The closed-loop system is stabilized using a discrete-time LQR controller. The discrete-time LQR is obtained by solving the infinite-horizon quadratic optimization problem
\begin{equation}
\centering
\scalebox{0.9}{$
\begin{aligned}
\min_{\{u_k\}} \quad & \sum_{k=0}^{\infty} \left( x_k^\top Q x_k + u_k^\top R u_k \right) \\
\text{s.t.} \quad & x_{k+1} = A_d \, x_k + B_d \, u_k,
\end{aligned}
$}
\label{eq:LQR_cost_discrete}
\end{equation}
where $Q$ and $R$ are the state and input penalty matrices, respectively, selected following Bryson's rule~\cite{joao_p_hespanha_linear_2010}, which normalizes each state and input by the inverse square of its maximum admissible deviation. The resulting weights were further adjusted to prioritize the regulation of selected states while balancing state performance and control effort. The penalty matrices are defined as
\begin{equation}
\label{eq:Q_R_matrices}
\centering
\scalebox{0.9}{$
\begin{aligned}
Q &= \mathrm{diag}\!\Big(
10\cdot\!\left(\tfrac{1}{0.1}\right)^{2},\;
5\cdot\!\left(\tfrac{1}{0.2}\right)^{2},\;
50\cdot\!\left(\tfrac{180}{10\pi}\right)^{2},\;
0.1\cdot\!\left(\tfrac{180}{90\pi}\right)^{2}
\Big), \\[3pt]
R &= 250\cdot\!\left(\tfrac{1}{2.5}\right)^{2}.
\end{aligned}
$}
\end{equation}
The resulting saturated state-feedback control law is given by
\begin{equation}
u_k = \mathrm{sat}\!\left(-K_{\text{LQR}} x_k\right),
\label{eq:LQR_law_discrete}
\end{equation}
where $\mathrm{sat}(\cdot)$ denotes the element-wise saturation operator enforcing the actuator voltage limits $u_k \in [-2.2,\, 2.2]\,\mathrm{V}$ and \( K_{\text{LQR}} = [-4.291,\ -8.142,\ -9.271,\ -0.574] \) is the corresponding optimal feedback gain.

\section{\uppercase{Region of Attraction Estimation}}
\label{sec:RoA_estimation}
This section explains the proposed methodology for computing a certified inner approximation of the RoA of the upright equilibrium. The approach starts from a certified positively invariant set, which provides a formal but conservative approximation of the RoA. In this work, a Lyapunov-based set is used as the initial certified invariant set. Building upon this initial invariant set, a predecessor-based implicit set expansion is employed to iteratively enlarge the certified region, yielding a significantly less conservative inner approximation under the nonlinear closed-loop dynamics. Finally, a RPI analysis is performed under bounded additive input disturbances, establishing robustness guarantees for the initial invariant set. 
\subsection{Lyapunov-Based Invariant Set}
\label{sec:Lyapunov-Based}
Following the Lyapunov-based construction presented in~\cite{fici_region_2026}, which is based on the proof of Theorem 4.7 in~\cite{khalil_nonlinear_2002}, the nonlinear closed-loop dynamics can be rewritten as
\begin{equation}
\label{eq:nonlinear_system_Lyap}
x_{k+1}=A_{\mathrm{cl}}x_k+g(x_k),
\end{equation}
where $A_{\mathrm{cl}}=A_d-B_dK_{\mathrm{LQR}}$ denotes the closed-loop linear dynamics and $g(x_k)$ collects the nonlinear terms. By Taylor's theorem, the nonlinear remainder is locally bounded as $|g(x_k)|<\gamma \, |x_k|$, for all states satisfying \(\|x_k\|<\rho\), where \(\rho>0\) defines a local neighborhood of the upright equilibrium. If \(\gamma\) satisfies
\begin{align}
\gamma <
\frac{-|PA_{\mathrm{cl}}|
+\sqrt{|PA_{\mathrm{cl}}|^2+\lambda_{\min}(Q_L)\lambda_{\max}(P)}}
{\lambda_{\max}(P)},
\end{align}
where P is the solution of the discrete-time Lyapunov equation $A_{\mathrm{cl}}^\top P A_{\mathrm{cl}}-P=-Q_L$, then the Lyapunov function
\(V(x_k)=x_k^\top Px_k\) decreases for all \(\|x_k\|<\rho\).

To obtain a positively invariant set, the largest Lyapunov sublevel set contained in this neighborhood is selected, yielding
\begin{align}
x_k^\top Px_k
\le
\lambda_{\min}(P)\rho^2.
\end{align}
This invariant set serves as the starting point for the predecessor-based expansion presented in the following subsection and as the basis for the RPI analysis.
\subsection{Predecessor-Based Expansion}
The Lyapunov-certified invariant set provides a conservative inner approximation of the RoA, as it only certifies states for which the Lyapunov function decreases directly. To enlarge this certified region while preserving convergence guarantees, a predecessor-based expansion is performed under the nonlinear closed-loop dynamics.

Let
\begin{equation}
\mathcal{V}_0=\left\{x:\;v_0(x)\leq1\right\},
\end{equation}
denote the initial Lyapunov-certified invariant set, where
\begin{equation}
v_0(x)=\frac{x^\top Px}{\lambda_{\min}(P)\rho^2}.
\end{equation}

The one-step predecessor of $\mathcal{V}_0$ is defined as
\begin{equation}
\mathcal{V}_1=f^{-1}(\mathcal{V}_0)
=\left\{x:\;f(x)\in\mathcal{V}_0\right\},
\end{equation}
where $f(\cdot)$ denotes the nonlinear closed-loop dynamics. Consequently, every state belonging to $\mathcal{V}_1$ reaches the invariant set $\mathcal{V}_0$ after one time step. Since $\mathcal{V}_0$ is positively invariant, every state in $\mathcal{V}_1$ is therefore guaranteed to converge to the upright equilibrium.

This construction can be applied recursively, yielding the sequence
\begin{equation}
\mathcal{V}_N=f^{-N}(\mathcal{V}_0),
\end{equation}
where $f^{-N}(\cdot)$ denotes the $N$-step predecessor operator. The resulting sets satisfy
\begin{equation}
\mathcal{V}_0
\subseteq
\mathcal{V}_1
\subseteq
\cdots
\subseteq
\mathcal{V}_N
\subseteq
\mathcal{ROA},
\end{equation}
where $\mathcal{ROA}$ denotes the region of attraction of the closed-loop system. The predecessor expansion is monotonic with respect to set inclusion and, as the number of predecessor iterations tends to infinity,
\begin{equation}
\bigcup_{N=0}^{\infty}\mathcal{V}_N=\mathcal{ROA},
\end{equation}
that is, the sequence of predecessor sets converges to the RoA. This property follows directly from the theory of backward reachable sets~\cite{serry_underapproximating_2025}.

Constructing the predecessor sets explicitly rapidly becomes analytically intractable due to the nonlinear dynamics. Instead, the predecessor expansion is represented implicitly. Since
\begin{equation}
\mathcal{V}_0
=
\left\{
x:\;
v_0(x)\le1
\right\},
\end{equation}
the one-step predecessor can be written as
\begin{equation}
\mathcal{V}_1
=
\left\{
x:\;
v_0(f(x))
\le1
\right\},
\end{equation}
and, by recursion,
\begin{equation}
\mathcal{V}_N
=
\left\{
x:\;
v_0\!\left(f^N(x)\right)
\le1
\right\},
\label{eq:implicit_predecessor}
\end{equation}
where $f^N(\cdot)$ denotes the $N$-step composition of the nonlinear closed-loop dynamics.

Equation~(\ref{eq:implicit_predecessor}) provides an implicit representation of the predecessor-expanded RoA approximation. Rather than explicitly constructing the boundary of $\mathcal{V}_N$, membership is determined by evaluating the implicit function
\begin{equation}
v_N(x)=v_0\!\left(f^N(x)\right).
\end{equation}
Consequently, a state belongs to the predecessor-expanded approximation if
\begin{equation}
v_N(x)\le1.
\end{equation}

Although the boundary of $\mathcal{V}_N$ does not generally admit a closed-form analytical description, the implicit representation enables exact membership tests.
\subsection{Robust Positive Invariance Analysis}
The RPI analysis of the Lyapunov-certified invariant set extends the nominal closed-loop dynamics introduced in (\ref{eq:nonlinear_system_Lyap}) by incorporating bounded additive disturbances acting at the control input. The disturbed closed-loop system is therefore written as
\begin{equation}
x_{k+1}=A_{\mathrm{cl}}x_k+g(x_k)+Bw_k+h(x_k,w_k),
\end{equation}
where $Bw_k$ denotes the linear contribution of the disturbance and $h(x_k,w_k)$ collects the higher-order nonlinear interaction terms introduced by the disturbance. The disturbance satisfies $\|w_k\|\le\bar{w}$.

Using the quadratic Lyapunov function
\[
V(x_k)=x_k^\top Px_k,
\]
obtained from the solution of the discrete-time Lyapunov equation in Section~\ref{sec:Lyapunov-Based}, the Lyapunov difference is given by
\[
\Delta V(x_k)=V(x_{k+1})-V(x_k).
\]
Unlike the nominal case, the additive disturbance introduces a constant contribution through the term $Bw_k$, which does not vanish at the equilibrium. Consequently,
\[
\Delta V(0)
=
(Bw_k)^\top P(Bw_k)
>
0,
\]
for any nonzero disturbance, implying that no neighborhood containing the origin can satisfy $\Delta V(x_k)<0$.

By combining the nonlinear remainder and the disturbance contribution into a single perturbation term, and applying norm inequalities together with the local Taylor bound on the nonlinear dynamics, a sufficient condition for $\Delta V(x_k)<0$ is obtained as
\begin{equation}
\label{eq:RPI}
p(\rho,\bar{w})<0,
\end{equation}
where $p(\rho,\bar{w})$ is a sixth-degree polynomial in the radius $\rho$, whose coefficients depend on the disturbance bound $\bar{w}$. In contrast to Section~\ref{sec:Lyapunov-Based}, the disturbance-dependent terms prevent the simplifications that reduce the sufficient negativity condition to a quadratic polynomial, resulting instead in a sixth-degree polynomial.

For a prescribed disturbance magnitude, the positive roots of $p(\rho,\bar{w})$ define the interval over which the Lyapunov function is guaranteed to decrease. Denoting these roots by $\rho_{\mathrm{in}}$ and $\rho_{\mathrm{out}}$, with
\[
0<\rho_{\mathrm{in}}<\rho_{\mathrm{out}},
\]
the decrease condition is certified for all states satisfying
\[
\rho_{\mathrm{in}}
<
\|x_k\|
\le
\rho_{\mathrm{out}}.
\]
Here, $\rho_{\mathrm{in}}$ characterizes the disturbance-dominated neighborhood around the equilibrium, while $\rho_{\mathrm{out}}$ determines the largest radius for which the dissipative effect of the nominal closed-loop dynamics dominates the disturbance contribution.

\section{\uppercase{Numerical and Experimental Results}}
\label{sec:numerical_results}
This section presents the numerical results obtained by applying the proposed methodology to the nonlinear TWIP. First, the Lyapunov-based invariant set is computed. Next, the predecessor-based expansion is performed to obtain an enlarged certified approximation of the RoA. Then the RPI condition is evaluated under bounded additive input disturbances. Finally, the resulting approximation is validated through Monte Carlo simulations and hardware experiments.

\begin{figure}[!t]
    \centering
    \includegraphics[width=\linewidth]{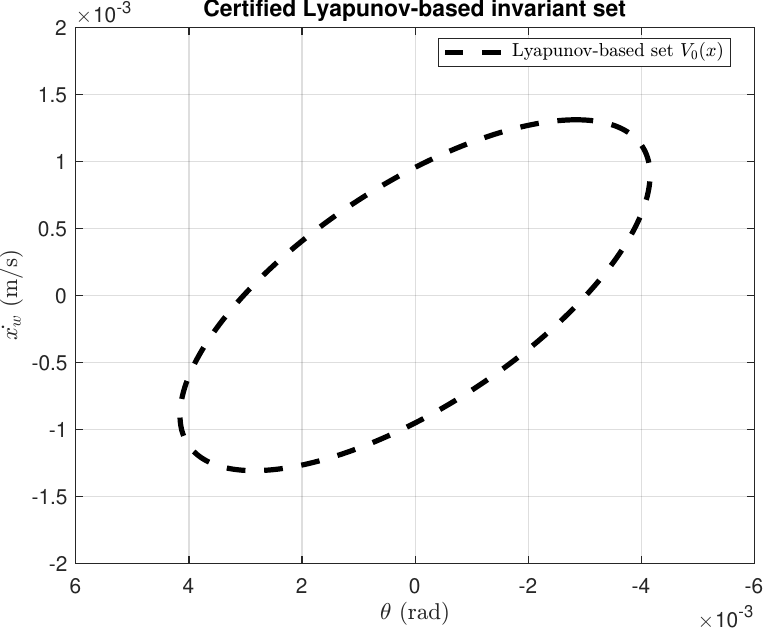}
    \caption{Certified Lyapunov-based invariant set.}
    \label{fig:lyapunov_set}
\end{figure}

Following the procedure described in Section~\ref{sec:Lyapunov-Based}, the quadratic Lyapunov function was obtained by solving the discrete-time Lyapunov equation with \(Q_L\) as the $4 \times4$ identity matrix, and the local bound on the nonlinear remainder was used to determine the largest neighborhood in which the Lyapunov decrease condition is satisfied. This yields a local validity radius of
\[
\rho = 2.724 \cdot 10^{-2}.
\]
The corresponding Lyapunov-certified invariant set is shown in Fig.~\ref{fig:lyapunov_set}, projected onto the $(\theta, \dot{x}_w)$ plane. As expected, the resulting set provides a certified inner approximation of the RoA around the upright equilibrium, although it is very conservative.

\begin{figure}[!t]
    \centering
    \includegraphics[width=\linewidth]{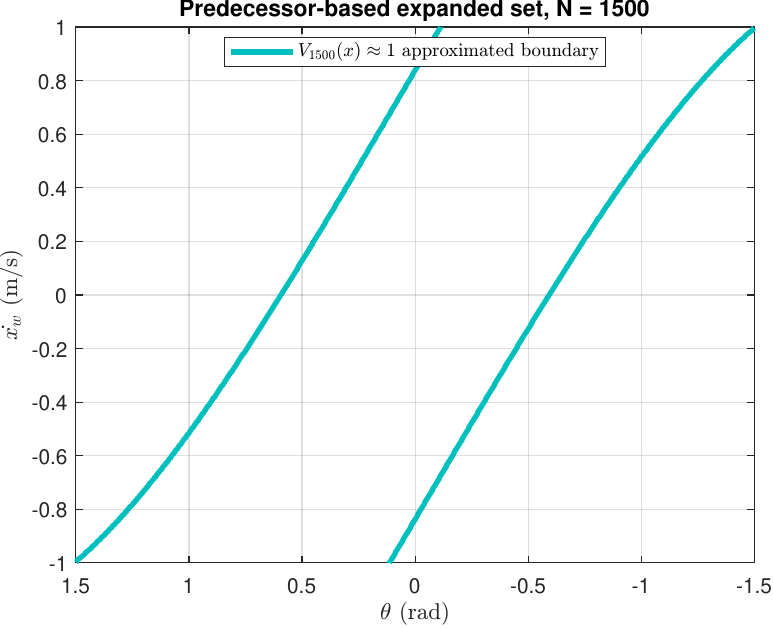}
    \caption{Implicit predecessor-expanded RoA. The contour obtained is an approximation.}
    \label{fig:predecessor_set}
\end{figure}

Starting from this certified invariant set, the predecessor-based expansion was computed over a prediction horizon of $N=1500$ sampling instants, corresponding to a simulation horizon of $15\,\mathrm{s}$. Since the predecessor-expanded set is defined implicitly through the condition $V_N(x)\le1,$ its boundary does not admit an explicit analytical representation. Therefore, the contour shown in Fig.~\ref{fig:predecessor_set} was approximated numerically solely for visualization purposes.

The approximation was obtained by evaluating the implicit predecessor function on a uniform $600\times600$ grid over the state subspace
\begin{equation}
\label{eq:MC_grid}
 \dot{x}_w\in[-1,1]~\mathrm{m/s},
 \qquad
 \theta\in[-1.5,1.5]~\mathrm{rad},   
\end{equation}
while fixing the remaining states to $x_w=0$ and $\dot{\theta}=0$. The boundary corresponding to $V_N(x)\approx1$ was then extracted by interpolation between neighboring grid points. It should be emphasized that only the plotted contour is approximated. The predecessor-expanded set itself remains exactly defined through the implicit membership condition $V_N(x)\le1$.

Finally, the sufficient condition~(\ref{eq:RPI}) was evaluated for a disturbance bound of $\bar{w}=7.5\times10^{-7}\,\mathrm{V}$. For this disturbance magnitude, the two positive roots of $p(\rho,\bar{w})$ are $\rho_{\mathrm{in}}=2.03\times10^{-3}$ and $\rho_{\mathrm{out}}=2.62\times10^{-2}$. Consequently, the Lyapunov decrease condition is satisfied for $2.03\times10^{-3}<\|x_k\|\leq2.62\times10^{-2}$, certifying robust positive invariance under the considered disturbance bound. This result provides a sufficient robustness certificate and does not represent the maximum disturbance magnitude that can be tolerated by the closed-loop system.

\subsection{Monte Carlo Validation}
Since the predecessor-expanded set is represented implicitly, its boundary has been approximated numerically for visualization. To assess the accuracy of this contour, a Monte Carlo validation was performed using the nonlinear closed-loop dynamics. The simulated trajectories provide an empirical reference for evaluating the agreement between the approximated boundary and the observed closed-loop behavior.

The validation was carried out by uniformly sampling 5000 initial conditions from the two-dimensional state subspace defined by $\dot{x}_w \in [-1,1]~\mathrm{m/s}$, and $\theta \in [-1.5,1.5]~\mathrm{rad}$ while fixing the position and the angular velocity to $x_w=0, \, \dot{\theta}=0 $ for visualization purposes. Each initial condition was propagated through the nonlinear closed-loop dynamics under the saturated LQR controller over a simulation horizon of $15\,\mathrm{s}$, matching the time horizon used to compute the predecessor expansion.

Each trajectory was classified according to whether it reached the Lyapunov-certified invariant set within the simulation horizon. Since this set is positively invariant, every trajectory entering it is mathematically guaranteed to converge to the upright equilibrium. The resulting stable and non-converging samples were then superimposed on the contour of the predecessor-expanded approximation defined by $V_N(x)=1$.

\begin{figure}[!t]
    \centering
    \includegraphics[width=\linewidth]{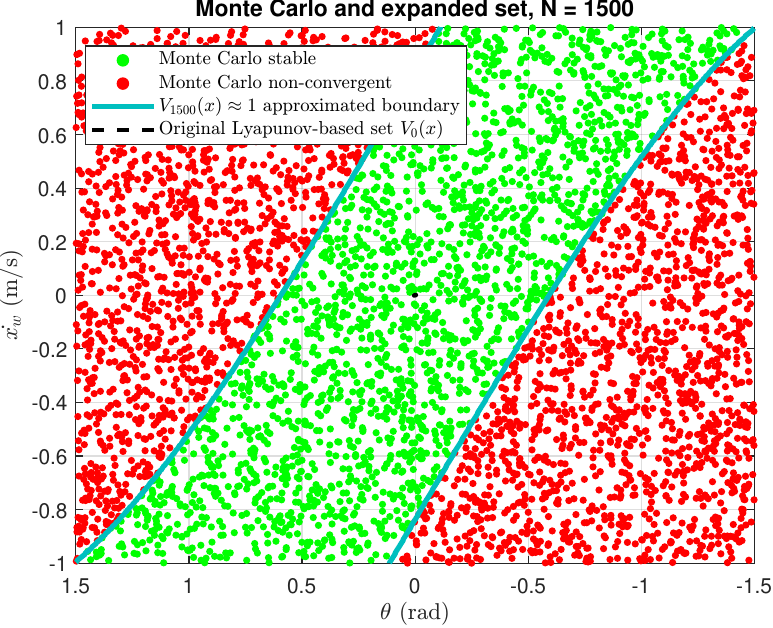}
    \caption{Monte Carlo validation of the predecessor-expanded RoA approximation. The initial Lyapunov-certified invariant set, visible as the small region near the origin, is shown for comparison.}
    \label{fig:combined_set}
\end{figure}

Figure~\ref{fig:combined_set} shows that the predecessor-expanded boundary closely follows the empirical separation between stable and non-convergent initial conditions. Compared with the initial Lyapunov-based invariant set, the predecessor expansion captures a substantially larger portion of the observed RoA while preserving formal stability guarantees. The close agreement between the approximated contour and the Monte Carlo classification supports the reliability of the approximated contour as a representation of the predecessor-expanded certified RoA.

\subsection{Experimental Validation}

To complement the certified RoA approximation, hardware experiments were conducted on the TWIP platform using the saturated LQR controller. Further details on the experimental implementation are provided in~\cite{detailleur_synthesis_2025}. The objective of these experiments is to assess whether the experimentally observed closed-loop behavior is consistent with the predecessor-expanded RoA approximation. In particular, for each disturbed state, the implicit predecessor function
\[
V_N(x)=V_0\!\left(f^N(x)\right),
\]
is evaluated. According to the proposed methodology, states satisfying $V_N(x)\le1$ belong to the certified predecessor-expanded approximation of the RoA, whereas states for which $V_N(x)>1$ lie outside the certified region. A video demonstration is provided.\footnote{\url{https://www.youtube.com/@GDLab-Research}}

Two representative experiments were considered. In the first experiment, the robot tracks a constant velocity reference of $-0.5\,\mathrm{m/s}$, together with the corresponding position reference. Short-duration input disturbances, applied as square pulses of duration $0.02\,\mathrm{s}$ with increasing amplitude up to $1.4\,\mathrm{V}$, drive the tracking error $e=x_{\mathrm{ref}}-x$ to
\[
e=
\begin{bmatrix}
-0.15&
-0.03&
-0.21&
1.50
\end{bmatrix}^{\!\top},
\]
where $x_{\mathrm{ref}}$ denotes the reference state corresponding to the prescribed constant velocity. Evaluating the implicit predecessor function yields
\[
V_{1500}(x)=5.3\times10^{-8}<1,
\]
confirming that the disturbed tracking error belongs to the predecessor-expanded RoA approximation. Consistent with this prediction, the robot successfully recovers and returns to the desired balancing configuration.

In the second experiment, the robot tracks a higher velocity reference of $-1.0\,\mathrm{m/s}$, while following the associated position trajectory. Under the same disturbance conditions, the tracking error reaches
\[
e=
\begin{bmatrix}
-0.40 &
-0.10 &
-0.50 &
-2.60
\end{bmatrix}^{\!\top}.
\]
In this case,
\[
V_{1500}(x)=1.0\times10^{8}>1,
\]
indicating that the tracking error lies outside the predecessor-expanded RoA approximation. Accordingly, the robot is unable to recover and loses stability.

These experimental results are fully consistent with the proposed predecessor-based RoA approximation. In both experiments, the implicit membership test correctly predicts whether the disturbed state belongs to the certified approximation, providing experimental validation of the proposed methodology.
\section{\uppercase{Conclusions}}
\label{sec:conclutions}
This paper investigated the computation of RoA approximations for a nonlinear TWIP by combining a certified Lyapunov-based invariant set with a predecessor-based implicit representation. The proposed approach significantly enlarges the initial certified region, yielding an RoA approximation that closely matches the empirically observed closed-loop behavior, as confirmed through Monte Carlo simulations and hardware experiments. In addition, the RPI analysis provides a formal robustness certificate for the initial Lyapunov-based invariant set under bounded additive input disturbances. 
While the predecessor-expanded region closely approximates the RoA, its implicit representation is not analytically tractable, requiring numerical procedures to approximate its boundary for visualization. Likewise, the robust positive invariance analysis provides guaranteed disturbance bounds, although the resulting certificates remain conservative with respect to the robustness observed in practice.

Future work will focus on deriving an explicit analytical representation of the predecessor-expanded region. In particular, one direction consists of computing the largest analytically describable set contained within the implicit predecessor representation, thereby obtaining an enlarged explicit certified inner approximation of the RoA. Alternatively, a certified branch-and-bound approach could be employed to verify the inclusion of candidate invariant sets, such as ellipsoids, within the implicit predecessor representation. Finally, future research will investigate less conservative approaches for robust invariance analysis, with the aim of obtaining less conservative robustness certificates that more accurately reflect the disturbance levels the system can tolerate in practice.
\section*{\uppercase{Acknowledgment}}
This work was financially sponsored by Franck Diard, Chief Software Architect at NVIDIA, through Fondation UniCA, and by 28 Digital.

\bibliographystyle{apalike}
\bibliography{references}
\end{document}